# Templated self-assembly of iron oxide nanoparticles

M. J. Benitez[1,2,*], P. Szary[1], D. Mishra[1], M. Feyen[2], A. H. Lu[2], O. Petracic[1,†] and H. Zabel[1]

[1]Institut für Experimentalphysik/Festkörperphysik, Ruhr-Universität Bochum, 44780 Bochum, Germany

[2]Max-Planck Institut für Kohlenforschung, 45470 Mülheim an der Ruhr, Germany

**Abstract:** We report on self-assembled iron oxide nanoparticle films on silicon substrates. In addition to homogeneously assembled layers, we fabricated patterned trenches of 40-1000 nm width using electron beam lithography for the investigation of assisted self-assembly. The nanoparticles with a diameter of 20 nm ± 7% were synthesized by thermal decomposition of iron oleate complexes in trioctylamine in presence of oleic acid. Samples with different track widths and nanoparticle concentration were characterized by scanning electron microscopy and by superconducting quantum interference device magnetometry.

**Keywords:** magnetic nanoparticles, self-assembly, templated self-assembly

## 1. Introduction

Magnetic nanoparticles (NPs) (often also termed nanocrystals or nanoclusters) still increasingly attract much interest, because they can be considered as building blocks for a multitude of applications, e.g. for high-density data storage media [1-4], spintronic devices [5, 6], photonic [7] or bio-medical systems [8-13]. In particular magnetite ($Fe_3O_4$) NPs are currently intensely discussed for medical applications due to their bio-compatibility and for spintronic devices because of their half-metallic properties [9, 13, 14].

A crucial factor especially for data storage media or spintronic devices is the quality of self-organization of NPs, which constitutes a popular bottom-up nanofabrication approach [1, 2, 15-17]. Often it is also desired to 'direct' or 'assist' the self-organization either because long-range order extending over large areas has to be achieved [1, 2] or because additional functionality needs to be included. To this end often a combination of self-organization of NPs and other fabrication techniques is employed [18, 19].

---

[*] E-mail: Maria.BenitezRomero@ruhr-uni-bochum.de
[†] E-mail: Oleg.Petracic@ruhr-uni-bochum.de

Here we report on self-assembled iron oxide NP arrays on silicon substrates. Furthermore, using electron beam lithography we fabricated patterned trenches of 40-1000nm width for templated self-assembly. The self-assembled NP arrays and the templated systems were characterized by superconducting quantum interference device (SQUID) magnetometry. We show that by combining 'top-down' and 'bottom-up' techniques, it is possible to obtain periodic patterned arrays of iron oxide NPs.

## 2. Experimental details

We have synthesized spherical iron oxide NPs with a mean diameter of 20 nm and a relatively narrow size distribution of 7% by thermal decomposition of iron oleate complexes in trioctylamine in presence of oleic acid as described by Park *et al* [20]. The NPs are coated with an oleic acid surfactant layer and are dissolved in toluene. This solution has then been used for the further processing steps.

For the first case of self-organized arrays the NPs were spin-coated on Si substrates with a native oxide. Typically approximately 0.2 ml of NP dispersion was spin-coated at 3000 rpm for 30 s and dried on a hot plate at 80°C for 20 min. Some samples were further annealed at 170° or 400°C in air.

For the second case of templated self-organization, periodic patterned lines were prepared by electron beam lithography with positive resist. As positive resist we used polymethyl meta acrylate (PMMA 200 K, 4%) spin-coated at 2000 rpm for 30 s to produce a ≈220 nm thick layer. Then, the PMMA layer was exposed to a focused 20-keV electron beam in a FEI

QUANTA FEG 200 scanning electron microscope (SEM) with a RAITH PC controlled pattern generator.

In order to fabricate parallel arrays of trenches of 130-1000 nm width, an electron dose of 400 $\mu C/cm^2$ and beam current of 71 pA was used. In the case of the smaller trenches, i.e. 40 nm width, an electron dose of 1000 pC/cm and a beam current of 51 pA was used. The exposed PMMA was then developed in a 1:3 mixture of methyl isobutyl ketone (MIBK) and isopropanol for 40 s, rinsed with isopropanol for 30 s and dried with a pure $N_2$ stream. The remaining resist from the developed area was removed using ion-milling. Furthermore, few nanometers were deliberately milled into the substrate for the assisted self-assembly of the NPs. Then NPs were spin-coated as in the case before. Finally, the resist parts were lift-off with acetone in an ultrasonic bath. This results in NPs inside the trenches only. The templated self-assembly is depicted schematically in Fig. 1.

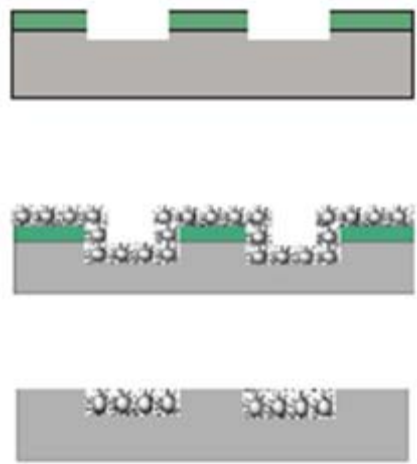

**Fig. 1.** Schematic of the templated self-assembly process. First, trenches were fabricated using electron beam lithography with subsequent ion-milling (top). Second, NPs are spin-coated (middle) and, third, resist is lift-off yielding NPs only in trenches (bottom).

Structural characterization of the NP films was carried out with the above mentioned SEM. Magnetization measurements were performed using a Quantum Design MPMS5 SQUID magnetometer. A detailed structural characterization of the composition of the NPs with the involved iron oxide phases after various heating protocols is described elsewhere [21].

## 3. Results and discussion

As a first step, self-organized monolayers and multilayers of NPs on top of Si substrates were studied. Fig. 2 shows an SEM image of a multilayer of NPs after spin-coating on a Si substrate and drying at 80°C for 20 min on a hot plate.

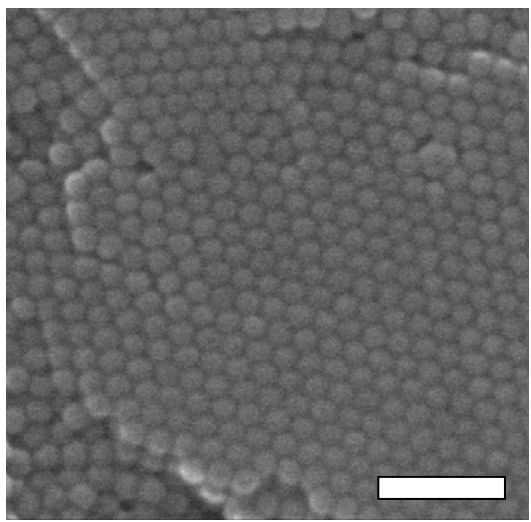

**Fig. 2.** SEM image of a self-organized multilayer of iron oxide NPs on top of an unstructured Si substrate. The NPs have a mean diameter of 20 nm and are coated with an oleic acid shell ($\approx 1$ nm). The bar corresponds to 100 nm.

One observes hexagonal close-packed ordering with grain sizes extending over regions of up to 300 nm. A previous study has shown that NPs prepared under these conditions yield NPs which contain two different stoichiometries and crystallographic structures, i.e. antiferromagnetic (AF) wüstite ($Fe_xO$) and ferrimagnetic maghemite ($\gamma$-$Fe_2O_3$) [21]. However, we find that annealing at 170°C in air yields predominantly maghemite NPs [21].

Fig. 3 shows zero-field cooled (ZFC) and field-cooled (FC) $M(T)$ curves of monolayer NP films after annealing in air at 80°C (a), 170°C (b) and 400°C (c). Panel (d) shows hysteresis loops for the case of annealing at 170°C.

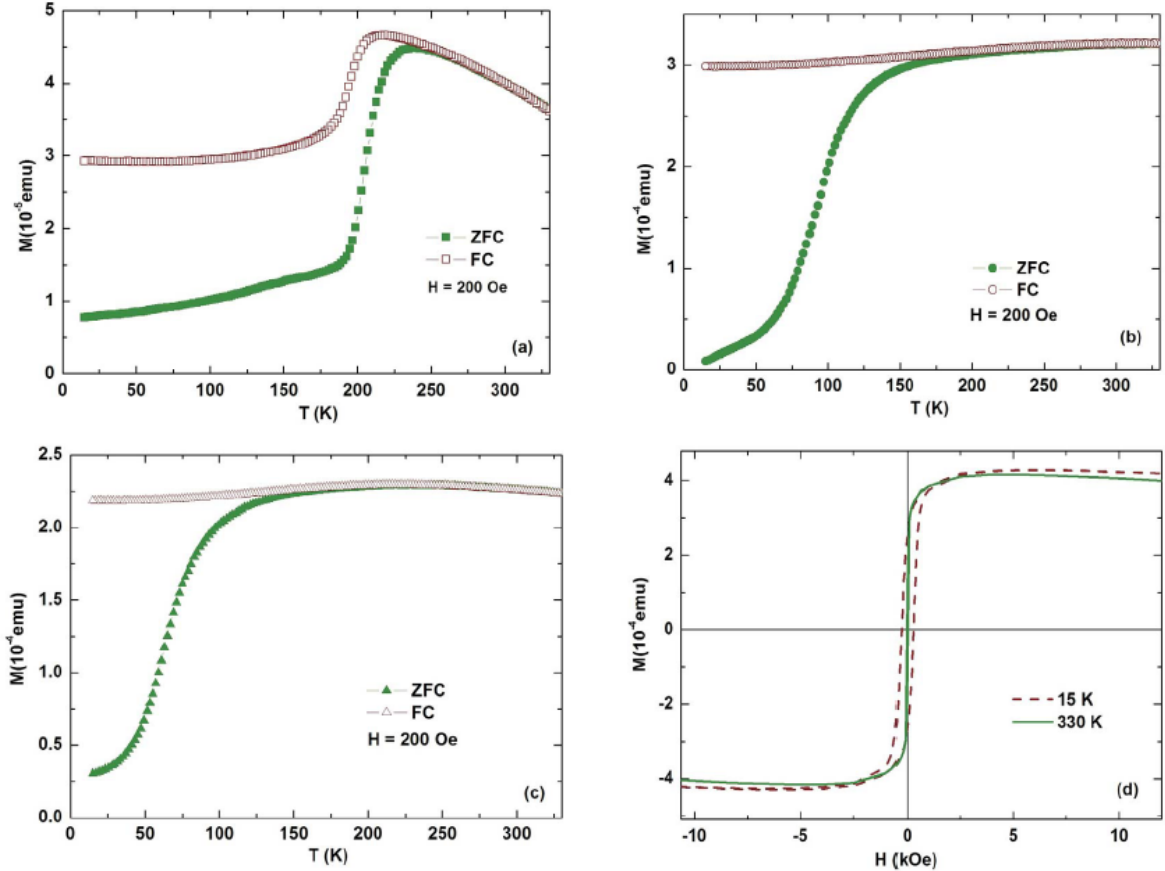

**Fig. 3.** Magnetization curves for a monolayer film of NPs annealed in air at different temperatures. *M vs. T* curves after ZFC (solid symbols) and after FC (open symbols) measured at 200 Oe for samples annealed in air at 80°C (a), 170°C (b) and 400°C (c). *M vs. H* hysteresis curves at 5 K after ZFC and after FC for sample annealed at 170°C (d).

Fig. 3 (a) reveals an atypical $M(T)$ behavior for the 80°C annealed case, i.e. a steep increase in the ZFC curve at ≈200 K, two peaks in the ZFC and FC curve at different temperatures and a strong decrease in the FC curve below the peak temperature. This behavior can be understood in terms of a superposition of two effects: Firstly superparamagnetic (SPM) behavior due to the ferrimagnetic maghemite iron oxide phase inside each NP and secondly exchange bias (EB) between the wüstite and the maghemite phase [21]. The EB breaks down near the Néel temperature of wüstite, i.e. $T_N = 198$ K and thus produces the steep increase or decrease in the ZFC and FC curve, respectively. The absence of a step or kink in the magnetization at 120 K, being a hallmark for the Verwey transition [22-24], indicates the absence of magnetite in the NPs.

Fig. 3 (b) and (c) show the magnetization curves of NPs annealed at 170°C and 400°C, respectively. Clearly a more regular SPM behavior is encountered here with a blocking temperature of $T_B$ = 300 K and 200 K, respectively. Obviously the NPs consist of mainly ferrimagnetic maghemite [21]. The slight decrease in the FC curve at low temperatures is likely due to inter-particle dipolar interactions [25-30]. In panel (d) is an example of the hysteresis loop shown for the case of the 170°C annealed system. One finds the expected unblocked or blocked SPM behavior at $T$ = 330 K or 15 K, respectively. The surprisingly square-like loop at 15 K could be due to a preferential anisotropy axis or due to superferromagnetic (SFM) collective inter-particle behavior [29, 30].

In a next step we investigated the magnetic properties of self-organized and templated NPs. The samples were dried at 80°C in air similar to the case above. Therefore, we expect again mixed-phase wüstite/maghemite NPs, showing a superposition of SPM and EB behavior.

Fig. 4 shows SEM images of samples with various trench widths, i.e. 1000 nm (a), 700 nm (b), 130 nm (c) and 40 nm (d). The samples having trench widths of 1000 nm and 700 nm exhibit grain sizes of about 200 – 300 nm, which is on the same scale as for the homogeneous films without trenches. The assisted self-assembly should become more effective for trench widths, which are smaller than the average grain size. This is indeed the case as can be seen in Fig. 4 (c) and (d) by the lack of voids in the direction parallel to the trenches.

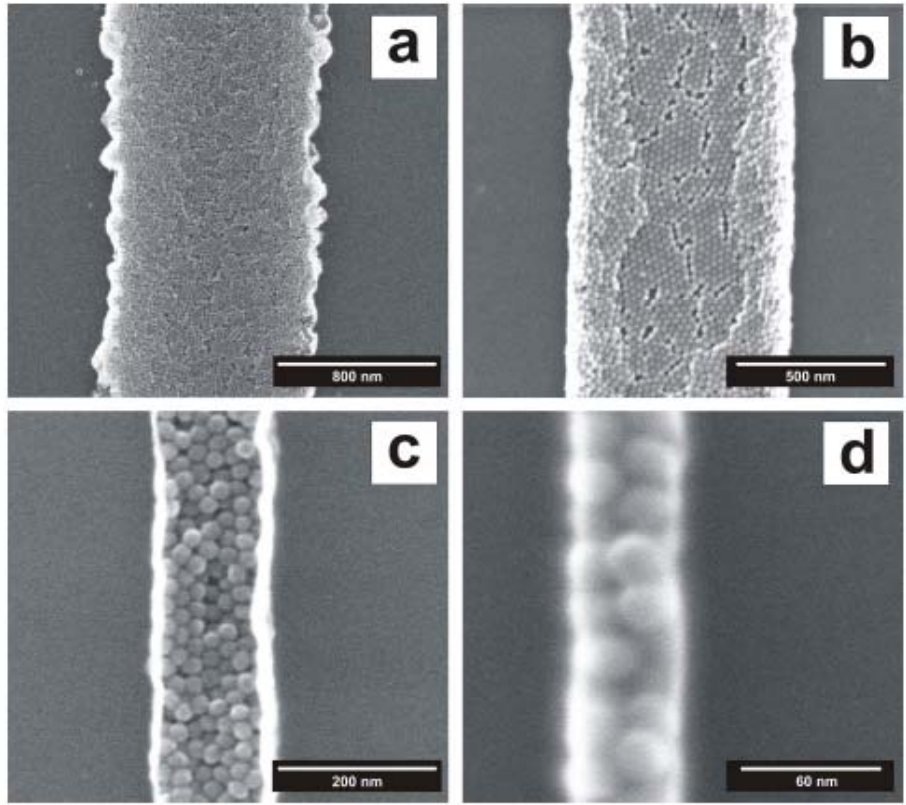

**Fig. 4.** SEM images of self-assembled NPs in patterned lines with different width 1000 nm (a), 700 nm (b), 130 nm (c) and 40 nm (d).

Magnetic characterization was performed on the patterned trenches of 130 nm width to study the influence of the confinement in the magnetic properties of the NPs. Fig. 5 (a) shows *M* vs. *T* curves after ZFC and after FC measured at 50 Oe and 200 Oe for NPs dried at 80°C. The magnetic field was applied parallel to the long axis of the trenches. One finds similar features to the one observed in the films, i.e., a peak in the ZFC and FC curves, irreversible behavior at low temperatures and a sudden increase in magnetization in the ZFC curve [21].

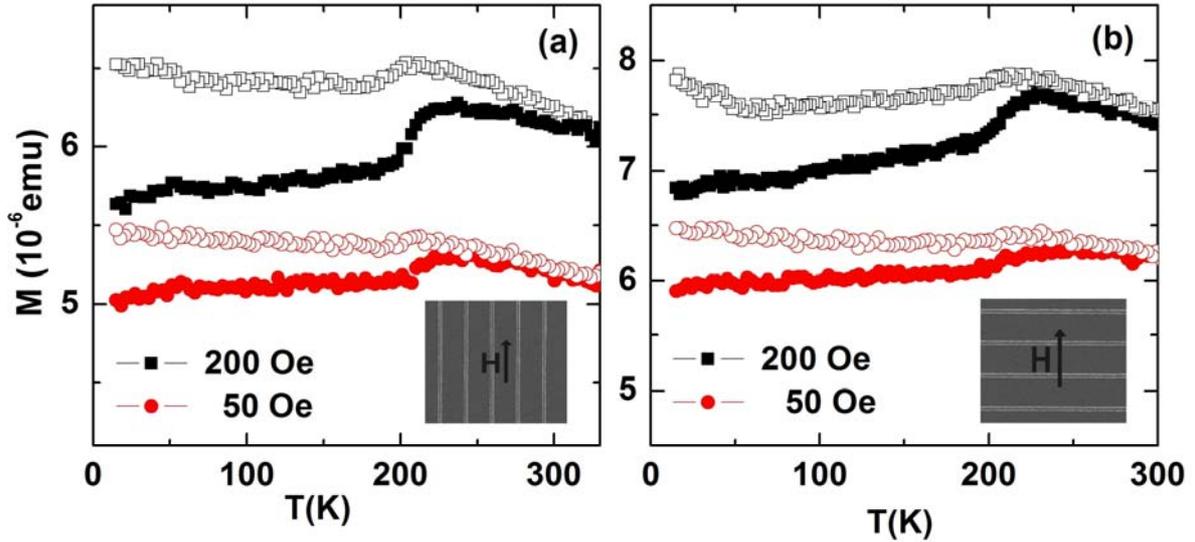

**Fig. 5.** $M$ vs. $T$ curves after ZFC (solid symbols) and FC (open symbols) measured at two applied fields, i.e. 50 Oe (circles) and 200 Oe (squares) for patterned trenches of 130 nm width, dried at 80°C. The magnetic field was applied along (a), and perpendicular (b) to the long axis of the trenches. The curves, ZFC and FC measured at 50 Oe are shown with an offset of $2.5 \times 10^{-6}$ emu (a) and with an offset of $4 \times 10^{-6}$ emu (b) for better clarity. The insets show the direction of the applied magnetic field.

Fig. 5 (b) shows $M(T)$ curves after ZFC and after FC measured at 50 Oe and 200 Oe for the case, where the magnetic field was applied perpendicular to the long axis of the trenches. The ZFC curve shows similar features to the ZFC curve measured when the magnetic field is applied parallel to the long axis. However, for the FC curve the magnetization is by about 20% higher than for the parallel direction and furthermore shows a new feature at low temperatures, i.e. an increase in magnetization below 50 K. From this we infer that the confinement introduces an anisotropy into the NP system, which might be due to a competition between field alignment, shape-anisotropy contributions, and exchange bias between wüstite and maghemite.

Magnetic characterization was also performed on the patterned trenches of 130 nm width after annealing at 170°C. In this case we again expect a predominant maghemite phase inside the NPs yielding SPM behavior [21]. The measurements do not show any influence of the direction of the applied magnetic field. Consequently the confinement of the NP array does not significantly affect the magnetization of NPs in the maghemite phase and at these structure sizes. Very likely effects can be observed for the smallest trench sizes of 40 nm.

Fig. 6 shows $M(T)$ curves after ZFC and after FC measured at 50 Oe and 200 Oe. In comparison to the NP films annealed at the same temperature, similar features were obtained.

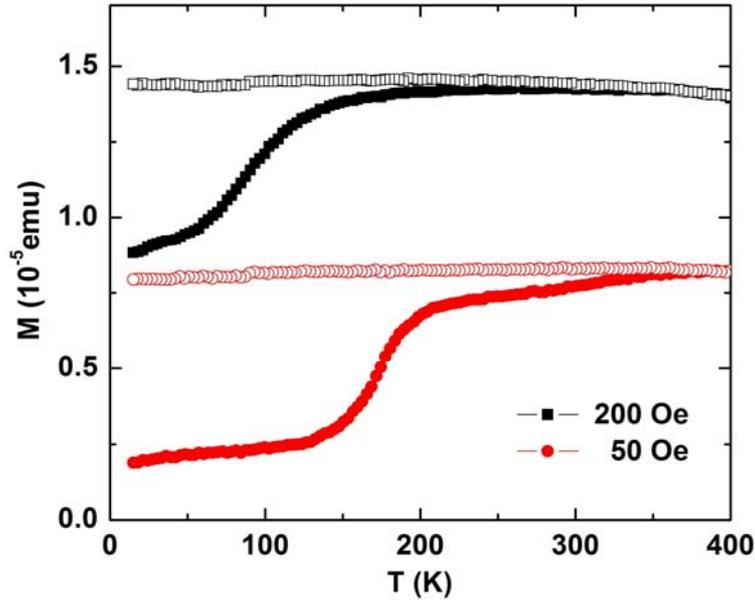

**Fig. 6.** $M(T)$ curves after ZFC (solid symbols) and FC (open symbols) measured at two applied fields, i.e. 50 Oe (circles) and 200 Oe (squares) for patterned trenches of 130 nm width annealed at 170°C. No influence of the field direction can be observed.

In summary, a simple approach for fabricating by e-beam lithography defined trenches with a width as low as 40 nm of iron oxide NPs was reported. Magnetometry studies were performed on trenches with widths of 130 nm. Samples annealed at 80°C in air, *viz.* consisting of wüstite/maghemite mixed-phase NPs show a weak influence due to the structuring. However, systems annealed at higher temperature containing mainly maghemite do not show modified behavior.

ACKNOWLEDGEMENTS

M.J.B. acknowledges support from the International Max-Planck Research School "SurMat" and D.M. from the NRW Graduate School: "Research with synchrotron radiation in the nano- and biosciences", Dortmund.